\documentclass[conference]{IEEEtran}
\IEEEoverridecommandlockouts
\usepackage{cite,subfigure}
\usepackage{amsmath,amssymb,amsfonts,amsthm}
\usepackage{algorithmic}
\usepackage{textcomp}
\usepackage{booktabs} 
\usepackage{makecell} 
\usepackage{xcolor}
\usepackage{pgf,pgfkeys,tikz,circuitikz,float,bm,caption}
\usepackage[ruled,linesnumbered,vlined]{algorithm2e}
\usepackage{graphicx,multirow}%
\usepackage{steinmetz}
\usepackage{siunitx,upgreek}
\def\BibTeX{{\rm B\kern-.05em{\sc i\kern-.025em b}\kern-.08em
    T\kern-.1667em\lower.7ex\hbox{E}\kern-.125emX}}

\graphicspath{{figures/}}

\begin{document}

\title{Range-Doppler Spoofing in OFDM Signals for Preventing Wireless Passive Emitter Tracking}

\author{\IEEEauthorblockN{Antonios Argyriou
	}
	\IEEEauthorblockA{Department of Electrical and Computer Engineering, University of Thessaly, Greece} 
}

\maketitle

\setlength{\abovedisplayskip}{6pt}
\setlength{\belowdisplayskip}{6pt}

\begin{abstract}
Passive emitter tracking (PET) algorithms can estimate both the range and Doppler of a wireless emitter when it uses orthogonal frequency division multiplexing (OFDM). In this paper we are interested to prevent this from happening by an unauthorized receiver (URx). To accomplish that we introduce in the transmitted signal a \textit{spoofing signal} that varies across subcarriers and successive OFDM symbols. With this technique the emitter is not only able to spoof its actual range and Doppler (allowing covert communication in terms of these two parameters), but is also capable of producing additional false emitter signatures to further confuse the URx. To evaluate the performance of our approach we calculate the range-Doppler response at the URx for different system configurations of an 802.11-based system.
\end{abstract}


\section{Introduction}
\label{section:introduction}
Wireless WiFi signals are omnipresent around us since they are emitted by most modern devices that are connected to a network. With passive emitter tracking (PET) techniques a receiver can use only the signal received directly from a wireless communications WiFi emitter (without any other reflection) and estimate fundamental parameters that characterize the behavior of the emitter. PET algorithms can estimate range/TDoA~\cite{spotfi15}, Doppler/AoA~\cite{cnf_2022_sam}, range/Doppler~\cite{Braun14}, the number of emitter antennas/AoA~\cite{cnf_2022_array}, etc. The same techniques can be used not only for tracking the emitter, but also for tracking non-emitting targets that are illuminated by a wireless source. This later concept is the basic principle behind \textit{passive RADAR}. 

\begin{figure}[!t]
	\centering
	\includegraphics[width=0.6\linewidth]{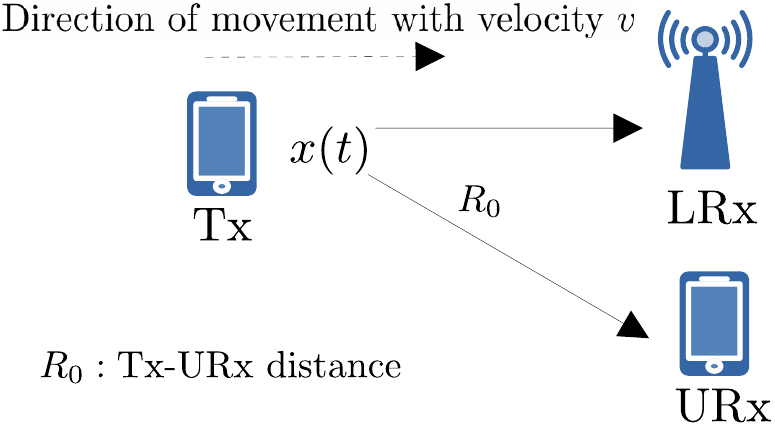}
	\caption{The transmitter constructs an OFDM signal $x(t)$ with a specific structure that we propose so that its spoofs its actual range and velocity $v$ relative to the unauthorized receiver (URx) that deploys PET algorithms. $x(t)$ is demodulated without performance degradation at the legitimate receiver (LRx).
	}
	\label{fig:system-ofdm-rd-spoofing}
\end{figure}

But despite the great appeal of these PET systems we argue that in some cases a wireless emitting device might not desire its tracking from an unauthorized receiver (URx) primarily out of privacy concerns. The question is then how can the emitter protect itself in such an environment. The question is not easy to answer but under certain conditions there are solutions, namely two: The first one is to have a third node transmit a jamming signal that disallows completely a URx to infer any parameter of the signal of interest from our emitter including Doppler or range~\cite{PhyCloak16}. The other solution is to embed a spoofing signal in the transmission so that it impairs the URx's ability to infer the correct Doppler (velocity) and range of the emitter. None of these two solutions should prevent a legitimate receiver (LRx) (Fig.~\ref{fig:system-ofdm-rd-spoofing}) from decoding the modulated symbols.

\textbf{The adversary:} In this paper we follow the second approach that is considerably more practical and we focus on WiFi orthogonal frequency division multiplexing (OFDM) signals. We first explore the capabilities of the URx, trying to consider a very sophisticated adversary equipped with OFDM RADAR signal processing. 
However, the basic OFDM RADAR algorithm that is used for estimating range and Doppler is applicable to monostatic OFDM RADAR where the transmitter and receiver of the waveform are one and the same~\cite{Braun14}. This means that the receiver that obtains an OFDM return signal can remove the known waveform and all the residual effects present in the signal are due to the target (its range, Doppler, etc.)~\cite{Braun14,cnf_2023_radarconf2}. But in this paper this is not directly applicable at the URx since our scenario is not active monostatic RADAR or even passive RADAR. Consequently, we modify the OFDM RADAR algorithm so that it works for PET of the range and Doppler by a URx. The core concept is the same: When the emitter uses OFDM, the URx leverages the signal across the subcarriers to estimate range and across symbols to estimate Doppler. But the URx now uses the known preamble of the wireless digital communication frame for removing it from the signal while it also demodulates the unknown data symbols also for removing them before estimating range and Doppler. This approach enables an OFDM RADAR algorithm for ranging and Doppler estimation to be applied to PET.

\textbf{The proposed spoofing strategy:} Our proposal is to alter in the transmitted OFDM signal the phase across the subcarriers and across OFDM symbols so that specific desired but incorrect velocity and range values are estimated at the URx that uses a PET algorithm. We essentially \emph{spoof} the range and Doppler of the emitter to point to a different desired value. 

Spoofing information in wireless communication signals has also been investigated in the literature, e.g. for synthetic aperture RADAR applications~\cite{Schuerger09}, and for frequency-modulated continuous wave (FMCW) RADAR~\cite{Komissarov21}. For PET in particular an experimental system for spoofing OFDM-based RADAR was presented in~\cite{Giusti18}. However, no systems have focused on range and Doppler spoofing of OFDM WiFi-based signals.

\section{Preliminaries \& Signal Model}

\subsection{PHY Preamble in 802.11 OFDM}
\begin{figure}[t]
	\centering
	\includegraphics[width=0.99\linewidth]{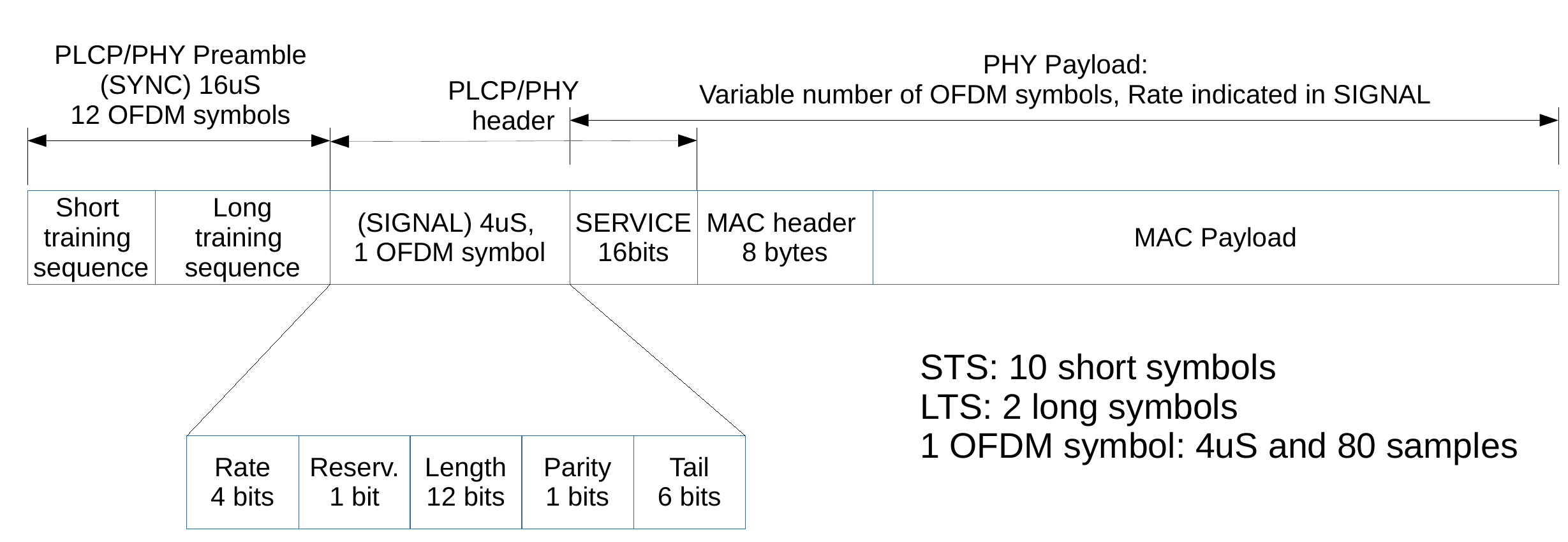}
	\caption{802.11a frame format.}
	\label{fig:wifi-frame-format}
\end{figure}

It is crucial to understand wireless frame transmission of the underlying wireless protocol, which in our case is 802.11 OFDM that was originally introduced in versions 802.11a/g. Each 802.11 frame consists of a preamble, a physical layer (PHY) header, and the MAC PDU (MPDU) which are the data from the perspective of the PHY (PHY payload).
The 802.11 preamble and the associated PHY header (Fig.~\ref{fig:wifi-frame-format}) are both transmitted at the lowest supported rate (BPSK using a channel coding rate 1/2 at 6 Mbps. The preamble is used by the receiver for frame detection, coarse frequency estimation, and timing synchronization. It consists of 10 copies of short OFDM symbols (\SI{0.8}{\micro\sec} each) that constitute the short training sequence (STS), and 2 long OFDM symbol that constitute the long training sequence (LTS). The STS and LTS have each a duration of \SI{8}{\micro\sec}. In terms of subcarrier utilization the STS is composed of 12 active subcarriers modulated by the elements of a fixed sequence~\cite{80211ac}. An important note is that in this paper besides the known STS and LTS preambles the URx demodulates the remaining OFDM data symbols for providing more data to the range-Doppler estimation algorithm.

\subsection{OFDM Wireles Network}
\label{section:topology}
As illustrated in Fig.~\ref{fig:system-ofdm-rd-spoofing} the transmitter (Tx) is part of the OFDM wireless communication network. There is a legitimate receiver (LRx) that is the intended receiver of the digital communication. Any WiFi-capable device can overhear the Tx frame transmissions and synchronize with their start. There is a URx that does exactly that but it does not have to be part of the network, i.e. decrypt the OFDM modulated symbols and bits. The URx has only to demodulate the OFDM signal and obtain the complex QAM symbols needed for the PET algorithm. The objective of the URx is to calculate the range-Doppler response for the Tx. Even though it is not necessary for our system to operate, for illustration purposes we assume that the Tx is moving towards the URx (from an initial distance $R_0$) with velocity $v$ and so the one-way delay of the signal at distance $R(t)$ is time-varying and equal to:\footnote{An AoA-dependent term in Doppler could also be included here but again is dropped for ease of exposition of our idea.}
\begin{align}
	\tau(t)=\frac{R(t)}{c}=\frac{R_0}{c}-\frac{vt}{c}
	\label{eqn:tau}
\end{align}

\subsection{OFDM RADAR Assumptions}
In the general case it might be difficult for a passive URx to estimate velocity and range from an OFDM signal unless the OFDM signal parameters are carefully tuned~\cite{Braun14}. However, under certain but not limiting assumptions it is possible. These are:
\begin{enumerate}
	\item The cyclic prefix (CP) duration is larger than the transmission time $\tau(t)$.
	\item The subcarrier distance is at least one order of magnitude larger than the largest occurring Doppler shift.
	\item If $f_c$ is the carrier frequency, then the Doppler shift is the same on every subcarrier that is $f_D=f_c\frac{v}{c}$ which is true for narrowband signals.
	\item The emitter distance remains constant during the transmission of 	one group of OFDM symbols that will be jointly processed (stop-n-hop assumption in RADAR signal modeling~\cite{book:fundamentals-of-radar-signal-processing}).
\end{enumerate}
The first and second assumptions are needed in order to prevent subcarrier de-othogonalization and allow demodulation of the received symbols~\cite{Braun14}. The second in particular is the most tricky to satisfy and depends on the standard. As an example for 802.11-based OFDM with a \SI{20}{\mega\hertz} channel the default subcarrier spacing is \SI{312.5}{\kilo\hertz} and with a $f_c$=\SI{5}{\giga\hertz} the Doppler is $f_D$=16.6$v$~\SI{}{\kilo\hertz} which is more than an order of magnitude smaller than \SI{312.5}{\kilo\hertz} for any realistic velocity $v$. For higher $f_c$ (e.g. in the \SI{24}{\giga\hertz} ISM band) better resolution could be achieved but the previous upper bound is reduced. Overall for typical OFDM systems in the ISM band passive algorithms are a plausible way to calculate range and Doppler.

\subsection{Signal Model}
It is important to re-state again that in this paper we consider a PET system, i.e. there is no signal reflection at all from a target. The "target" is the emitter of the digital communication signal. Hence, we consider standard channel models for wireless communication which, as we will soon see, still embed Doppler/velocity, and path loss due to range, and exclude aspects that are not applicable (e.g. there is no illuminated target so no need to model a cross-section).

Regarding the channel model we assume a dominant line-of-sight (LOS) path between the transmitter and the receiver that is modeled by a Rician statistical channel model. The Doppler shift is $f_D=f_c\frac{v}{c}$. If the total signal power is $P_r$ the complex channel gain coefficient is given by
\begin{equation}
	s(t)=\sqrt{\frac{KP_r}{K+1}}e^{j2\pi f_Dt+\phi}+\sqrt{\frac{P_r}{K+1}}g(t), \label{eqn:rician-channel-gain}
\end{equation}
where $K$ is the Rician factor that corresponds to the ratio of the power $\frac{K}{K+1}P_r$ of the LOS path versus the aggregate power $\frac{P_r}{K+1}$ from the remaining paths, and $g(t)$ is the complex Gaussian random process that corresponds to Rayleigh fading. $P_r$ is calculated based on the Friis path loss formula. In the high $K$ regime \eqref{eqn:rician-channel-gain} reduces to only the first LOS component. Note however, that $\phi$ is not deterministic but also a random variable (RV) uniformly distributed in $[0,2\pi]$, since it is also the result of scattering~\cite{ozdogan19}. For easier presentation we define $h$=$\sqrt{P_r}e^{\phi}$ that captures the impact of the channel excluding Doppler in~\eqref{eqn:rician-channel-gain}. 

Based on the previous discussion, the baseband continuous time signal model for transmission in a flat fading channel at the URx becomes:
\begin{align}
y(t)=hx(t-\tau(t))e^{-j2\pi f_c\tau(t)}+w(t)
\label{eqn:signal-model-analog}
\end{align}
With the narrowband assumption and~\eqref{eqn:tau} we have $x(t-\tau(t))=x(t-\frac{R_0}{c}+\frac{v}{c}t)\approx x(t-\frac{R_0}{c})$. 
In the above $x(t)$ represents the modulated signal, and $w(t)$ is the AWGN sample. In our model we ignore the sampling clock offset (SCO) even though we are interested in the correct demodulation of a symbol (by sampling when the matched filter output peaks), and also the carrier frequency offset (CFO). A more detailed passband and baseband signal model that could extend the above and includes CFO can be found in~\cite{cnf_2023_radarconf2}. 

We can expand the previous model that includes the impact of Doppler when $x(t)$ is the result of wideband multi-carrier modulation and in our case OFDM. In this case the previous channel model because it considers only the LOS path offers flat fading. With $N$ subcarriers that can contain data, pilot symbols, or a combination of both (depending on the standard), the desired OFDM symbol in continuous time is:
\begin{equation}
	x(t)=\frac{1}{\sqrt{N}}\sum_{k=0}^{N-1}X[k]e^{j2\pi f_k t},~~0\leq t \leq T_N.
	\label{eqn:ct-ofdm} 
\end{equation}
$X[k]$ is the complex symbol modulated onto subcarrier $k$ at frequency $f_k=k\Delta f$, $\Delta f$ is the subcarrier spacing, and $T_N=\frac{N}{\Delta f}$ is the OFDM symbol duration. By sampling this at times $t=n/f_s$, and recalling that for the fraction of the subcarrier frequency relative to the sampling rate it is $f_k/f_s=k/N$, we get a digital frequency $k/N$ and the discrete form of the OFDM signal at the transmitter:
\begin{equation}
	x[n]=\frac{1}{\sqrt{N}}\sum_{k=0}^{N-1}X[k]e^{j2\pi n k/N},~~0\leq n \leq N-1
	\label{eqn:idft-ofdm} 
\end{equation}
This is effectively the IDFT of $X[k]$ allowing thus its well known hardware-efficient implementation. 

We now replace~\eqref{eqn:ct-ofdm} into \eqref{eqn:signal-model-analog} and take into account the one-way delay $\tau(t)$ in~\eqref{eqn:tau}. We have that:
\begin{align}
	&y(t)=\frac{h}{\sqrt{N}}\sum_{k=0}^{N-1}X[k]e^{j2\pi f_k( t-\frac{R_0}{c})} e^{-j2\pi f_c(\frac{R_0}{c}-\frac{vt}{c})}+w[n] \nonumber	
\end{align}
This introduces a phase shift in the baseband model that depends on each subcarrier namely $e^{-2\pi f_k\frac{R_0}{c}}$ and a time varying phase shift that depends on the Doppler frequency $f_D=\frac{v}{c}f_c$ namely $e^{2\pi f_c\frac{vt}{c}}$ since we already assume the same Doppler regardless of the subcarrier index $k$.\footnote{In the wideband case the Doppler shift is not the same for each subcarrier and the signal model we develop here changes substantially as discussed in~\cite{cnf_2023_radarconf2}.} We also set $h^{'}$=$he^{-2\pi f_c\frac{R_0}{c}}$. We sample $y(t)$ at $t$=$n/f_s+mT_N$ to get the fast and slow time data cube and notice that the RHS of the last expression becomes:
\begin{align}
	&	\frac{h^{'}}{\sqrt{N}}\sum_{k=0}^{N-1}X[k]e^{j2\pi f_k( \frac{n}{f_s}+mT_N-\frac{R_0}{c})} e^{j2\pi f_D(\frac{n}{f_s}+mT_N)}+w[n] \nonumber	
\end{align}
This is simplified further by noticing that $e^{j2\pi f_k mT_N}=1$, since $f_kT_N$=$k\Delta f T_N$=$kN$. We also assume that the Doppler within one OFDM symbol is negligible so $e^{-j2\pi f_D\frac{n}{f_s}} \approx 1$ and we have:
\begin{align}
y[n,m]	=\frac{h^{'}}{\sqrt{N}}\sum_{k=0}^{N-1}X[k]e^{j2\pi f_k( \frac{n}{f_s}-\frac{R_0}{c})} e^{j2\pi f_DmT_N}+w[n],	\label{eqn:reflected-signal-model-discrete-1}
\end{align}
This expression gives the final signal model for the $m$-th OFDM symbol of duration $T_N$.

\section{Spoofing and Range-Doppler Generation}

\subsection{Range-Doppler Spoofing}
The discussion in the last section considered the nominal operation of the emitter that transmits an OFDM signal and the corresponding received signal at the URx, while the same signal model will dictate reception at the LRx (albeit with a different channel gain). Now we discuss what the emitter does with the proposed spoofing strategy. In our system the emitter takes the $N$ frequency-domain (FD) QAM symbols $X[k]$ and does a simple mixing/multiplication with the spoofing signal. We set the baseband spoofing signal in the frequency domain to be
\begin{align}
U[k,m]=e^{-j2\pi f_k\frac{R_\text{sp}}{c}} e^{j2\pi f_\text{sp}mT_N},
\label{eqn:spoofing-signal-fd}
\end{align}
where $R_\text{sp}$ indicates the spoofed range and $f_\text{sp}$ the spoofed Doppler frequency. This signal depends on the subcarrier $k$ and the OFDM symbol $m$. The baseband signal that is actually transmitted for the $m$-th symbol is $U[k,m]X[k]$.\footnote{We drop the index $m$ from $X[k]$ that should indicate in which OFDM symbol $m$ the $k$-th QAM symbol belongs to.}

\subsection{The URx with the PET OFDM RADAR Algorithm}
We know that when the signal is received at the OFDM receiver (URx in our scenario), the passive OFDM RADAR algorithm should process it with a DFT as discussed in~\cite{Braun14,cnf_2023_radarconf2}. Hence, the URx synchronizes to the start of the first OFDM symbol until the expected number of OFDM symbols is received. Upon downconversion and digitization, DFT is performed. After the DFT of~\eqref{eqn:reflected-signal-model-discrete-1} is calculated for a complete OFDM symbol of $N$ samples (it has also been affected by \eqref{eqn:spoofing-signal-fd}), we can sample for each $X[k]$ of the $m$-th OFDM symbol after $T_N$, i.e. at time $t=T_N=n/f_s \rightarrow n=T_Nf_s$. So we have that the DFT of the $m$-th OFDM symbol is
\begin{align}
	\tilde{Y}[k,m]&=h^{'} X[k]e^{-j2\pi f_k\frac{(R_0+R_\text{sp})}{c}} e^{j2\pi(f_D+f_\text{sp})mT_N}\nonumber\\
	&+W[k,m],0\leq k \leq N-1,0\leq m \leq M-1.	\label{eqn:signal-model-discrete-pre-process}
\end{align}

Upon downconversion and digitization, the URx also demodulates the OFDM symbols denoted as $\hat{X}[k]$ based on the signal in~\eqref{eqn:reflected-signal-model-discrete-1}. However, for this part we use the Moose algorithm~\cite{moose94} for removing frequency/phase offsets since it is an established well-known principle for synchronization in OFDM receivers. 
After demodulation our algorithm divides~\eqref{eqn:signal-model-discrete-pre-process} by $\hat{X}[k]$ to remove QAM modulation for the $m$-th symbol:
\begin{align}
	\frac{\tilde{Y}[k,m]}{\hat{X}[k]}&= 
	h^{'} \frac{X[k]}{\hat{X}[k]} e^{-j2\pi k\Delta f\frac{(R_0+R_\text{sp})}{c}} e^{j2\pi(f_D+f_\text{sp})mT_N}\nonumber\\
	&+\frac{W[k,m]}{\hat{X}[k]},0\leq k \leq N-1,0\leq m \leq M-1.	\label{eqn:signal-model-discrete-pre-process2}
\end{align}
When the demodulation performance is high then $\hat{X}[k]$ is more accurately estimated which means that \eqref{eqn:signal-model-discrete-pre-process2} will contain only the range and Doppler effects. Note that when errors accumulate then the URx cannot estimate with high quality neither the real nor the spoofed range-Doppler response since it is impaired by the ratio $X[k]/\hat{X}[k]$ that introduces random phase shifts for each QAM symbol $k$. 

Finally, the OFDM-based passive RADAR algorithm at the URx performs 2D DFT on the signal in~\eqref{eqn:signal-model-discrete-pre-process2} across the $k,m$ indexes with sampling periods $\Delta f$ and $T_N$ to obtain the range-Doppler response. The peak in the rane-Doppler response will occur at position $(R_0+R_\text{sp})/c$ and $f_D+f_\text{sp}$. Again, the reader is referred to~\cite{Braun14,cnf_2023_radarconf2} for more details regarding the range-Doppler response derivation in OFDM RADAR and for the wideband case in particular in~\cite{cnf_2023_radarconf2}.

\subsection{False Emitter Generation}
The transmitter can also generate a subset of the transmitted OFDM symbols with a specific set of spoofed range-Doppler parameters and a subsequent set with different ones (e.g. with a periodicity of a few OFDM symbols) according to~\eqref{eqn:signal-model-discrete-pre-process2}. The resulting signal model at the URx is also that in \eqref{eqn:signal-model-discrete-pre-process2}. The question is which set of OFDM symbols will the URx use in order to generate the range-Doppler response with the PET algorithm we described in the last subsection. This is something that is generally unknown and random but we evaluate its impact in our simulations.

\begin{figure}[t]
	\centering
	\includegraphics[width=0.95\linewidth]{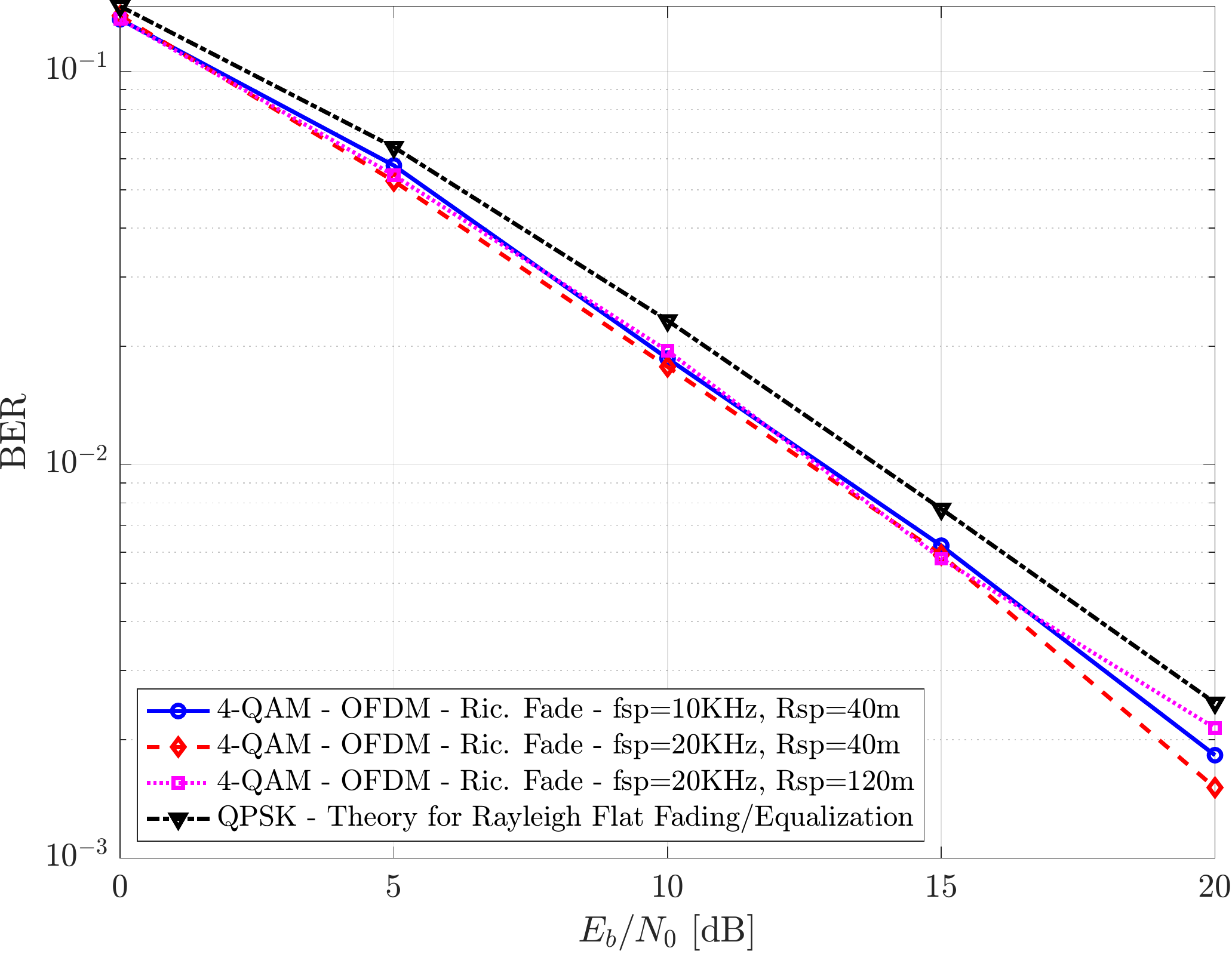}
	\caption{Bit error rate (BER) results for different values of the spoofed Doppler frequency $f_\text{sp}$ and range $R_\text{sp}$.}
	\label{fig:ber-ofdm}
\end{figure}

\subsection{LRx Demodulation} Inserting artificial Doppler creates inter-carrier interference (ICI) in the OFDM signal which is not desirable for the LRx. The question then is then since orthogonality is destroyed on purpose can we recover it at the receiver? The answer is yes if we insert a frequency shift that is recoverable by the receiver synchronization algorithm. Similar schemes that insert artificial frequency shifts have also been investigated in the literature but for conveying additional data~\cite{Wang18}. In 802.11 tolerances of up to \SI{625}{\kilo\hertz} can be accommodated by the frequency offset estimators~\cite{80211ac}. The same is true also for non-OFDM single-carrier continuous wave signals where an artificial frequency variation can be recovered in that case with a phased-lock loop (PLL)~\cite{jnl_2020_phy}.

\section{Simulations}
The objectives of our evaluation are twofold: First to measure the performance of the LRx in terms of BER and verify that the joint range-Doppler spoofing does not affect it. Second, produce range-Doppler responses at the URx and verify how much it is close to the desired spoofed Doppler and range. We considered an 802.11-based OFDM system of 64 subcarriers in a \SI{20}{\mega\Hz} channel, out of which 52 contained data, 4 subcarriers were used for pilot signals, while the remaining 8 were empty according to the standard. The URx processes a number of $M$=50 consecutive OFDM symbols for producing the range-Doppler response. The signal-to-noise ratio (SNR) refers to the power of the useful signal in our model vs the power of the noise at the URx receiver.

\textbf{BER Performance:} Fig.~\ref{fig:ber-ofdm} presents the BER performance of the LRx over the Rician fading channel when it uses the Moose algorithm~\cite{moose94} which uses established well-known principles for synchronization in OFDM receivers. The LRx has no reason to deploy the joint Doppler estimation algorithm developed in this paper since it only needs to find the overall phase shift and remove it for demodulation. Our results indicate that this standard algorithm is robust to different frequency ($f_\text{sp}$) and range ($R_\text{sp}$) spoofing values that do not affect its performance at the LRx.

\begin{figure}[t]
	\centering
	\subfigure[URx SNR of \SI{10}{\dB}. The lower SNR results in decreased demodulation performance for $\hat{X}$.]{\includegraphics[width=0.99\linewidth]{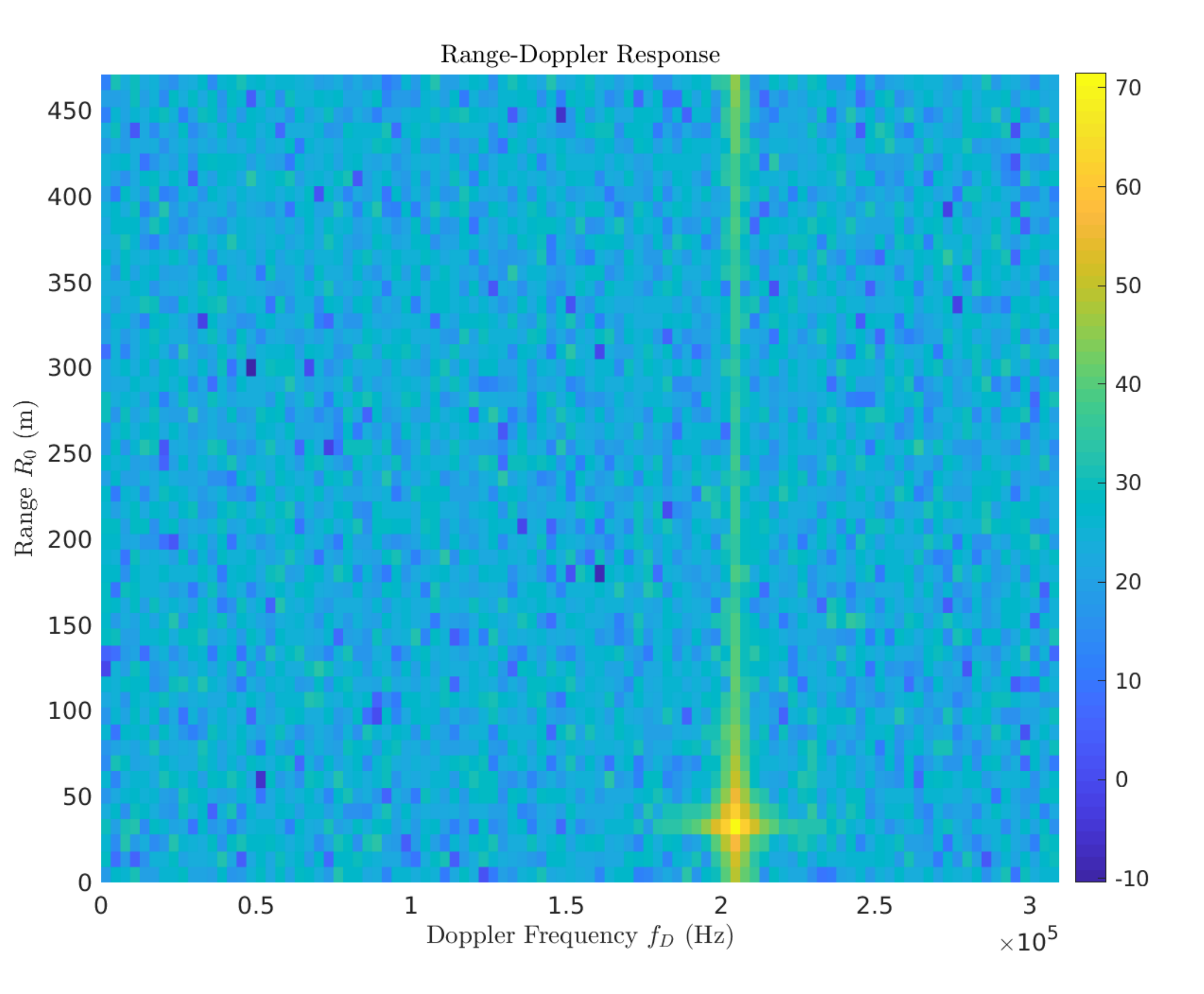}
	}\hspace{-0.18cm}
	\subfigure[URx SNR of \SI{30}{\dB}. SNR increase results in better demodulation performance for $\hat{X}$ and improved quality signature.]{\includegraphics[width=0.99\linewidth]{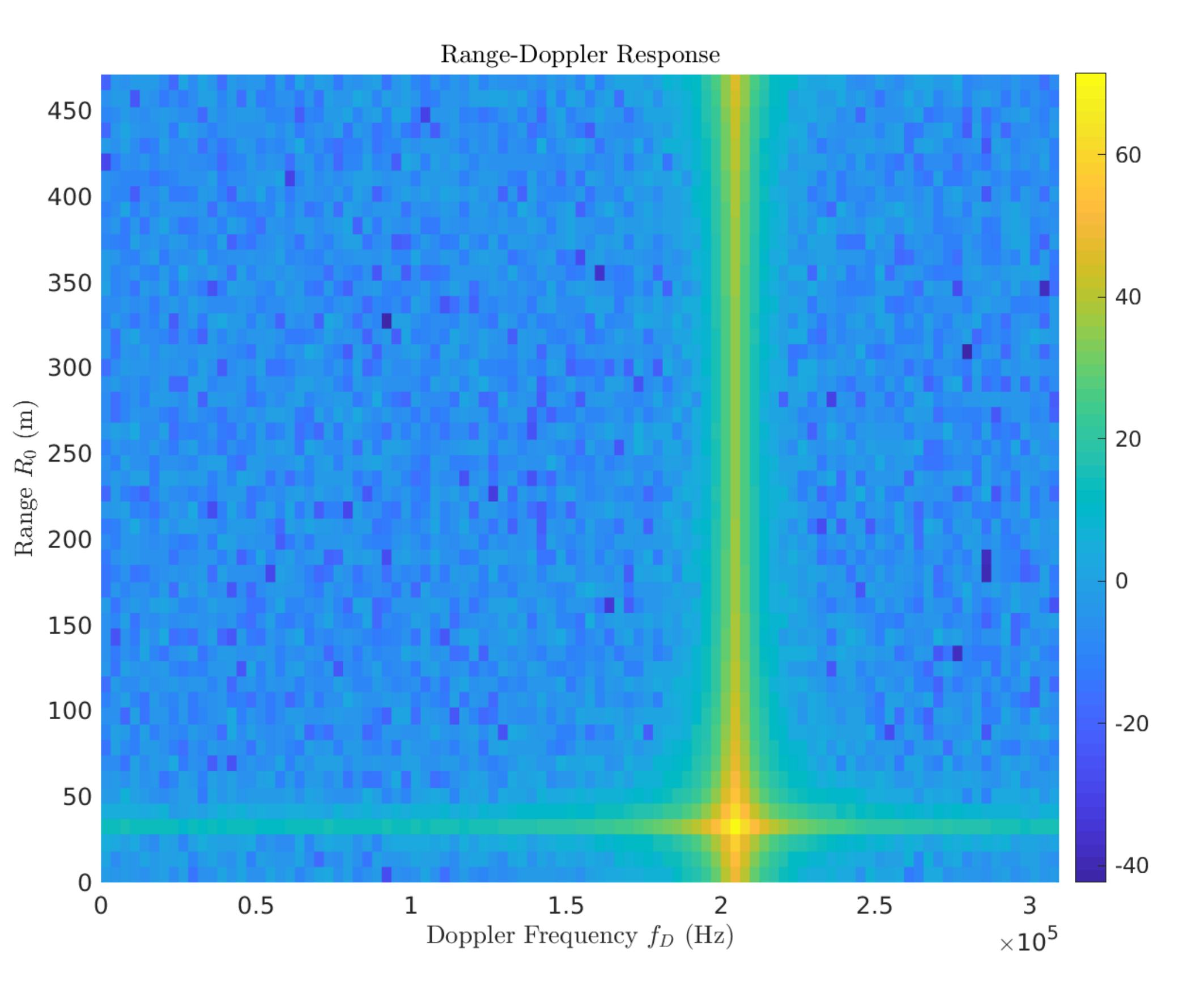}
	}
	\caption{Range-Doppler responses for different SNR at the URx (and different $\hat{X}$ demodulation performance).}
	\label{fig:radar-ofdm-range-doppler1}
\end{figure}

\textbf{Range-Doppler at URx:} In Fig.~\ref{fig:radar-ofdm-range-doppler1}(a) we present the range-Doppler response for an SNR of \SI{10}{\dB} that leads to worse BER performance for $\hat{X}[k]$ and so poor quality of range-Doppler responses. The random values that the ratio $\hat{X}[k]/X[k]$ takes introduces random phase shifts all over the range-Doppler response diagram. The range-Doppler response quality is significantly improved when SNR is improved (Fig.~\ref{fig:radar-ofdm-range-doppler1}(b)). Higher SNR leads to improved demodulation which means that $\hat{X}[k]/X[k]$ tends to 1 most of the time.

\begin{figure}[t]
	\centering	
	\subfigure[20\% of OFDM symbols used at the URx correspond to the primary false emitter. The secondary false emitter is more pronounced.]{\includegraphics[width=0.99\linewidth]{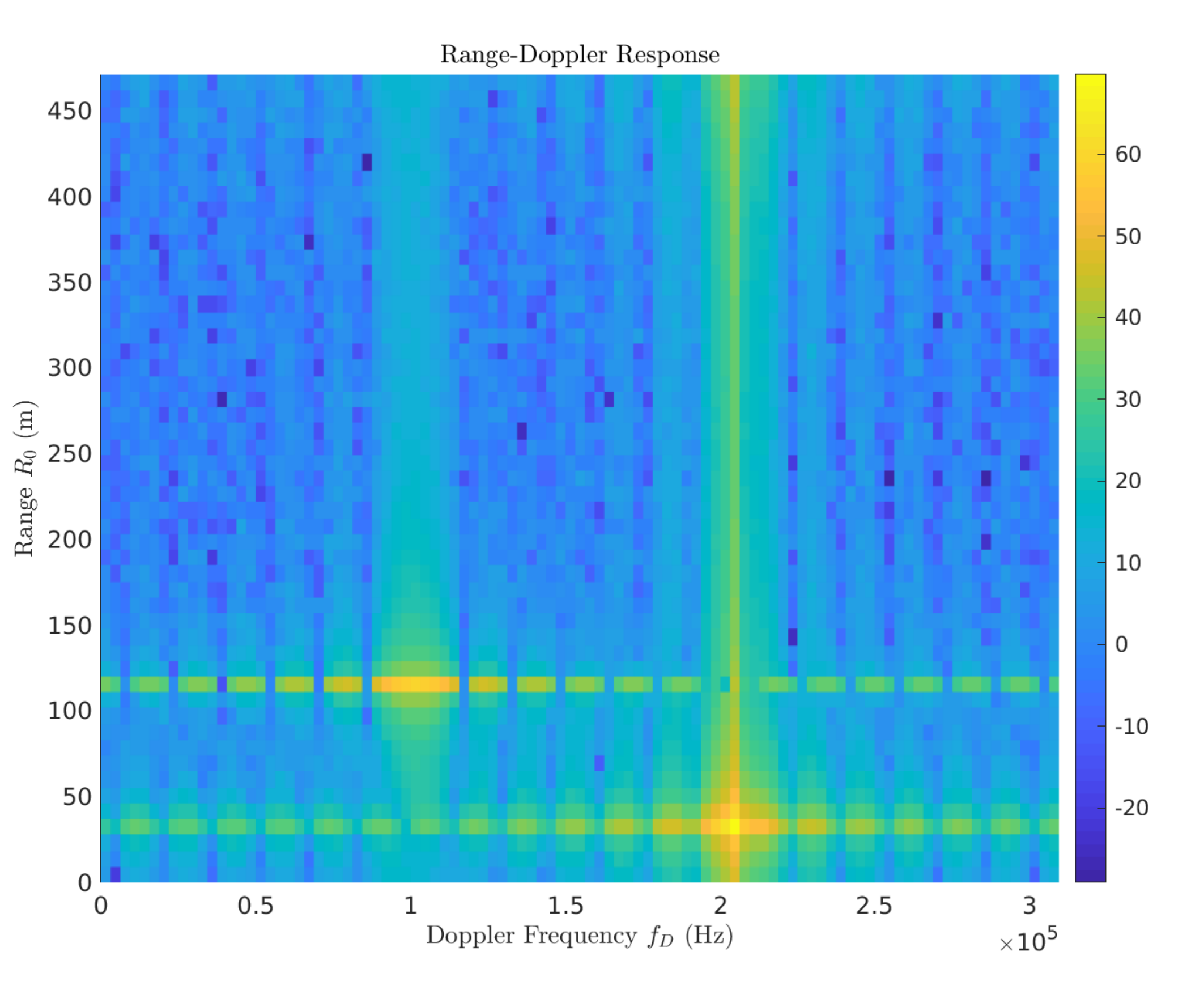}				\label{radar-ofdm-range-doppler_10outof50overlap}
	}\hspace{-0.18cm}
	\subfigure[40\% of OFDM symbols used at the URx correspond to the primary false emitter. The primary false emitter signature is improving.]{\includegraphics[width=0.99\linewidth]{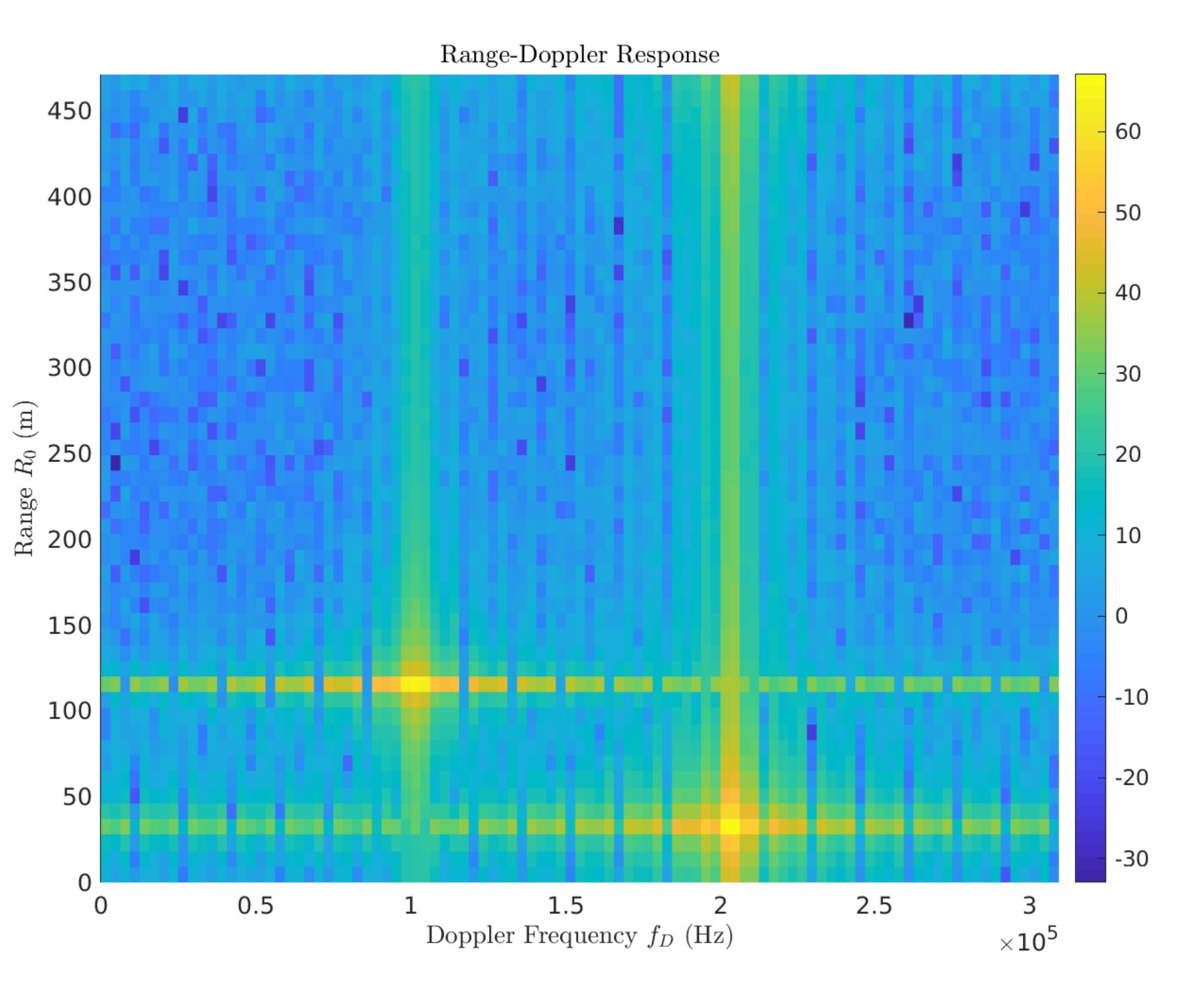}
		\label{fig:radar-ofdm-range-doppler_20outof50overlap}
	}
	\caption{Range-Doppler responses for different fraction of OFDM symbols spoofing different emitters.}
	\label{fig:radar-ofdm-range-doppler2}
\end{figure}

\textbf{Range-Doppler at URx and false emitter generation:} To evaluate the ability for produce false signatures consider that the actual emitter generates range-Doppler signatures for two false emitters. To do that it uses \eqref{eqn:spoofing-signal-fd} and two different sets of values for $f_\text{SP}$, $R_\text{SP}$ so that one subset of the $M$ OFDM symbols corresponds the first emitter, and the next subset to the second emitter. In these results the spoofed Doppler for the primary emitter is \SI{10}{\kilo\Hz} and the spoofed range \SI{120}{\m}. We name the emitter that we used in the paragraph before at \SI{20}{\kilo\Hz} and \SI{40}{\m} as the secondary false emitter. For calculating the range-Doppler response the URx uses all the $M$ OFDM symbols as defined previously.
In Fig.~\ref{radar-ofdm-range-doppler_10outof50overlap}, \ref{fig:radar-ofdm-range-doppler_20outof50overlap} we present results for different ratio of spoofed OFDM symbols that were used at the Tx (from a total of $M$=50), and we notice that as this fraction is increased the specific false emitter is experiencing a clearer signature. This indicates that the resulting false signature depends on the randomly selected OFDM symbols. In any case the range-Doppler response has the desired form in terms of the spoofed range and Doppler values. 

\section{Conclusions}
In this paper we presented a new approach for improving the privacy of an OFDM wireless communication emitter by preventing range and Doppler (velocity) estimation by an unauthorized receiver that uses recently established OFDM-based RADAR algorithms. The basic idea suggests the insertion of an artificial (spoofed) Doppler frequency and range in the transmitted signal that depends on the subcarrier and the specific OFDM symbol. This disallows the range-Doppler estimation algorithm to produce correct estimates of the channel-induced Doppler and range. Additional false emitters can be generated for disorienting further the URx. The final result is a scheme that is more robust to information leaks from OFDM emitters without compromising demodulation performance at the desired receiver. In the future we plan investigate a channel where besides the LOS component there is strong multipath that might allow passive coherent localization (PCL) algorithms to be used by the URx.

\bibliographystyle{IEEEtran}
\bibliography{../../../../tony-bib}
\end{document}